\documentclass{article}

\usepackage[preprint]{neurips_2025} 

\workshoptitle{Shallow Networks Are All You Need: RBF-PIELMs for Fast and Interpretable PDE Learning}

\usepackage[utf8]{inputenc} 
\usepackage[T1]{fontenc}    
\usepackage{hyperref}       
\usepackage{url}            
\usepackage{booktabs}       
\usepackage{amsfonts}       
\usepackage{nicefrac}       
\usepackage{microtype}      
\usepackage{xcolor}         
\usepackage{caption}
\usepackage{subfig}
\usepackage{booktabs} 
\usepackage{graphicx}
\usepackage{subcaption}
\usepackage{amsmath}
\usepackage[inkscapeformat=pdf]{svg}
\title{Deep vs. Shallow: Benchmarking Physics-Informed Neural Architectures on the Biharmonic Equation}

\author{%
 Akshay Govind Srinivasan \\
  Department of Mechanical Engineering\\
  Indian Institute of Technology, Madras\\
  \texttt{me22b102@smail.iitm.ac.in} \\
  \And
  Vikas Dwivedi \\
 Creatis Biomedical Imaging Laboratory\\
   INSA-Lyon, France\\
  \texttt{vikas.dwivedi@creatis.insa-lyon.fr} \\
  \And
  Balaji Srinivasan \\
   Department of Data Science and AI \\
   Indian Institute of Technology, Madras \\
  \texttt{sbalaji@iitm.ac.in}  \\
}

\begin{document}

\maketitle

\begin{abstract}
Partial differential equation (PDE) solvers are fundamental to engineering simulation. Classical mesh-based approaches (finite difference/volume/element) are fast and accurate on high-quality meshes but struggle with higher-order operators and complex, hard-to-mesh geometries. Recently developed physics-informed neural networks (PINNs) and their variants are mesh-free and flexible, yet compute-intensive and often less accurate. This paper systematically benchmarks RBF-PIELM, a rapid PINN variant—an extreme learning machine with radial-basis activations—for higher-order PDEs. RBF-PIELM replaces PINNs' time-consuming gradient descent with a single-shot least-squares solve. We test RBF-PIELM on the fourth-order biharmonic equation using two benchmarks: lid-driven cavity flow (streamfunction formulation) and a manufactured oscillatory solution. Our results show up to \(350\times\) faster training than PINNs and over \(10\times\) fewer parameters for comparable solution accuracy. Despite surpassing PINNs, RBF-PIELM still lags mature mesh-based solvers and its accuracy degrades on highly oscillatory solutions, highlighting remaining challenges for practical deployment. We open-source our code for reproducibility and future extensions\footnote{The link to the Anonymous Repository can be found \href{https://anonymous.4open.science/r/RBF-PIELM-553F/}{here}}  
\end{abstract}

\section{Introduction}
Numerical solution of ODEs and PDEs underpins modeling in fluid and solid mechanics. Classical discretization methods—finite difference (FDM) \cite{strikwerda2004finite}, finite element (FEM) \cite{hughes2012finite}, and finite volume (FVM) \cite{versteeg2007finite}—offer high accuracy but incur significant computational cost due to their larger stencils, and their reliance on expensive mesh generation in complex or large-scale domains~\cite{Yagawa2011-cr}. This motivates a need for developing mesh-free methods for solving ODEs and PDEs for complex domain.
Physics-Informed Neural Networks (PINNs) \cite{raissi2019physics} provide a mesh-free alternative by embedding governing equations into neural network training. While they have demonstrated success across a range of applications, their training times are often significantly higher than those of state-of-the-art numerical solvers for achieving comparable accuracy~\cite{McGreivy2024}. Hence making them inefficient as compared to traditional solvers. Moreover, challenges remain in terms of sensitivity to hyperparameter choices and the lack of interpretability in their learned representations~\cite{krishnapriyan2021characterizing}.

\citeauthor{dwivedi2019physics} introduced Physics-Informed Extreme Learning Machines (PIELMs) as an efficient alternative to PINNs, which merge the physics-based loss of PINNs with the shallow architecture of Extreme Learning Machines (ELMs) \cite{huang2006extreme}. Unlike gradient-based training \cite{shalev2017failures}, ELMs fix input weights randomly and compute output weights via a pseudo-inverse, yielding orders-of-magnitude faster training while retaining universal approximation guarantees \cite{huang2006extreme,huang2006universal}. PIELMs thus provide rapid and data-efficient PDE solvers \cite{dwivedi2020physics,xu2022bayesian}. Since PIELMs initialize their input weights randomly, the resulting hidden features lack physical interpretability and exhibit limited alignment with the underlying physics, hence cannot be initialized in a physics-informed manner.

~\citeauthor{DWIVEDI2025130924} introduced Radial Basis Function-based PIELMs (RBF-PIELMs), which extend PIELMs by replacing random hidden features with localized Radial Basis Functions (RBFs) \cite{buhmann2003radial,schaback2007kernel}. RBFs offer interpretable, physics-aware activations: each hidden unit corresponds to a localized ``receptive field,'' and centers can be aligned with domain geometry, boundary layers, or measurement data \cite{kansa1990multiquadrics}. Their widths can be tuned to resolve sharp gradients such as shear layers or shocks. Moreover, RBF networks retain universal approximation guarantees \cite{Park1991}, preserving theoretical expressivity while maintaining the efficiency of PIELMs through a closed-form solution.

\paragraph{Contributions}  
This paper investigate the applicability of RBF-PIELMs as a fast, interpretable, and flexible solver for linear PDEs. Specifically:
\begin{itemize}
\item We investigate the applicability of RBF-PIELMs with geometry-aware initialization for solving higher order PDE.
\item We benchmark against PINNs on the lid-driven cavity problem, showing similar accuracy with much lower training cost and model complexity.
\item We demonstrate expressivity of RBF-PIELM by solving a complex biharmonic problem~\cite{PAN2025114254} and evaluate it using the Method of Manufactured Solutions (MMS).
\end{itemize}\vspace{-0.5em}

The remainder of this paper is organized as follows: 
Section~\ref{sec:rbf-pielm-math} presents the mathematical formulation of RBF-PIELM; Section~\ref{sec:test-lid-driven} presents the lid-driven cavity benchmark presented in \citeauthor{marchietal}; Section~\ref{sec:test-mms} introduces the MMS study presented in~\citeauthor{PAN2025114254};Section~\ref{sec:conclusion} reports and discusses results and concludes with future directions.
\section{Methodology: RBF-PIELM}
\label{sec:rbf-pielm-math}
Let $u:\Omega\subset\mathbb{R}^m\to\mathbb{R}^n$ be a function that satisfies Equation~\ref{eqn:linear-op} where $\mathcal{L}(u)(x)$ is a linear differential operator 
in a given domain $\Omega$
\begin{equation}
\label{eqn:linear-op}
    \mathcal{L}(u)(x)+f(x)=0,\quad x\in\Omega,
\end{equation}
with boundary operator $\mathcal{B}$ and data $g$ on $\partial\Omega$. In RBF-PIELM, the function approximation is:
\begin{equation}
\label{eqn:fnapprox}
    \hat{u}(x)=\sum_{i=1}^{N^*} c_i\,\phi_i(x),
    \qquad
    \phi_i(x)=\exp\!\left(-\tfrac{\|x-x_i\|^2}{2\sigma_i^2}\right),
\end{equation}
where $x_i$ and $\sigma_i$ denote the center and width of the $i$-th RBF \cite{buhmann2003radial}. Centers and widths may be chosen randomly or in a problem-aware like choosing more RBFs near walls and boundary layers, or they can be initialized in a data-driven manner \cite{dwivedi2025kerneladaptivepielmsforwardinverse}. Given that the interior collocation points are represented as $\{x_j^{\Omega}\}$ and boundary points as $\{x_k^{\partial}\}$, the residuals are
\begin{align}
    \mathcal{R}_{\Omega}(x) &= \mathcal{L}(\hat{u})(x)+f(x), \quad x\in\Omega,\\
    \mathcal{R}_{\partial}(x) &= \mathcal{B}(\hat{u})(x)-g(x), \quad x\in\partial\Omega.
\end{align}
\vspace{-1em}

Enforcing residuals at collocation points yields an over-constrained linear system (assuming that $N^* < |\Omega| + |\partial \Omega| $) as in Equation~\ref{eqn:linear-system} with entries of $A$ defined by the PDE and boundary operators. 
\begin{equation}
\label{eqn:linear-system}
    A\,c=b,
\end{equation}
The coefficients $c$ are then obtained via the penrose pseudo-inverse. These coefficients can then be substituted in Equation~\ref{eqn:fnapprox} to obtain approximate solution to Equation~\ref{eqn:linear-op}.
\vspace{-1.5em}
\section{Numerical Experiments}
\subsection{Biharmonic Equation with Smooth Solution: Lid-Driven Cavity Flow}
\label{sec:test-lid-driven}
The lid-driven cavity problem is a classical benchmark in incompressible flow simulations, and in this work we adopt its stream function–vorticity (biharmonic) formulation. The biharmonic equation arises in modeling mixing in cavities, micromechanical flows, and in solid mechanics through the Airy stress function~\cite{radice2021airy_biharmonic}. Its fourth-order nature makes it challenging to solve using mesh-free methods due to the need for accurately enforcing multiple boundary conditions and handling higher derivatives. We use it both for its ubiquity and difficulty, benchmarking our method by comparing centerline velocities with the results of~\citeauthor{marchietal}.
 The velocity $\mathbf{u}=(u,v)^T$ satisfy
\begin{equation}
  -\nabla^2 \mathbf{u} + \nabla p = 0,\quad \nabla \cdot \mathbf{u} = 0.
\end{equation}
Introducing the stream function $\psi(x,y)$ via
 $ u=\tfrac{\partial \psi}{\partial y}$, $\quad v=-\tfrac{\partial \psi}{\partial x}$, reduces the equations to the biharmonic form as in Equation~\ref{eqn:biharmonic}.
\begin{equation}
\label{eqn:biharmonic}
    \psi_{xxxx} + 2\psi_{xxyy} + \psi_{yyyy} = 0, \quad (x,y)\in[0,1]^2,
\end{equation}
with no-slip boundary conditions. The lid motion is enforced by $\psi_y(x,1)=1$, while all other walls remain fixed. The definitions of Boundary Conditions is given in Appendix Section~\ref{sec:BCLidDriven}.

\begin{figure}[htbp!]
    \centering
\subfloat[\centering Horizontal velocity \(u (y)\) along \(x=1/2\).]{%
        \label{uCenterProfile}\includegraphics[width=0.48\textwidth]{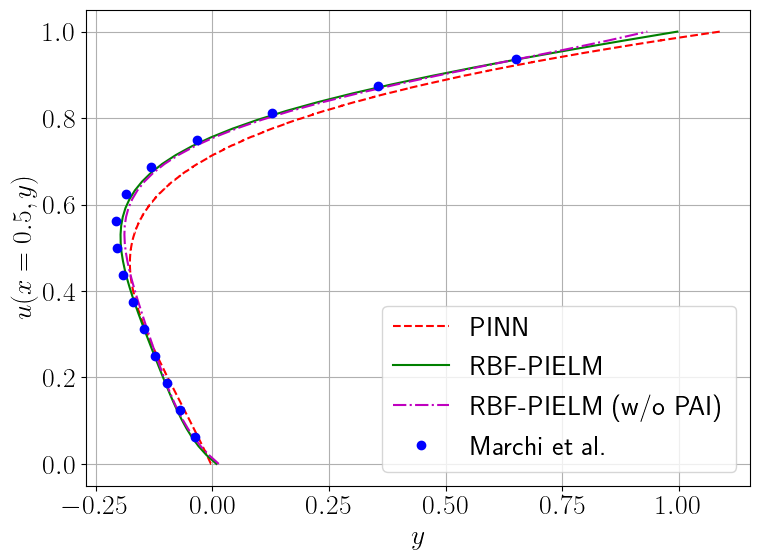}
    }
    \hfill
    \subfloat[\centering Vertical velocity \(v (x)\) along \(y=1/2\)]{%
        \label{vCenterProfile}\includegraphics[width=0.48\textwidth]{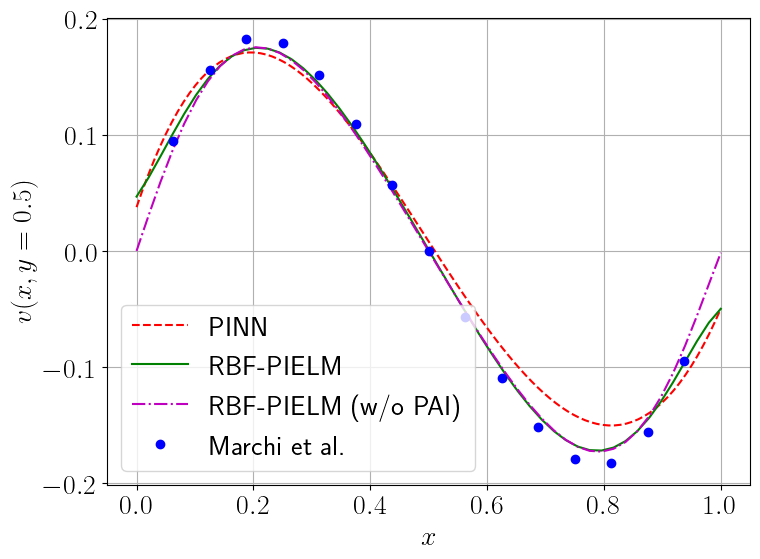}
    }
    \caption{ Centerline Velocity Profiles: RBF-PIELM vs. PINNs. RBF-PIELM attains better accuracy while training \(350\times\) faster than PINNs.}
    \label{fig:pielm-flow-results}
\end{figure}

\paragraph{Results and Discussion}
\begin{itemize}
    \item \textbf{Physics-Aware Initialization} To accurately capture the sharp velocity gradients near the cavity walls, the collocation points are distributed with Chebyshev spacing (Refer Appendix Section ~\ref{sec:chebyshev}) and the RBF Kernels are over-sampled near boundaries, ensuring higher resolution near boundaries. The RBF standard deviation is chosen using a heuristic: $\sigma = 0.3 + 0.93\left({L_{\min}\over L_{\max}}\right)$, where $L_{\min}$ is the distance of the kernel center to the nearest wall and $L_{\max}$ the maximum possible distance in the domain. Further details about tuning the parameters of the heuristics is discussed in Appendix Section~\ref{sec:hyper-param}. This physically motivated choice ensures narrower kernels near boundaries and broader ones in the interior, providing more physics-aware initialization of input layer parameters. 
    \item \textbf{PIELM vs PINN} Figure~\ref{uCenterProfile} and ~\ref{vCenterProfile} compares the centerline velocities obtained from PINN, RBF-PIELM and \citeauthor{marchietal}. We observe that RBF-PIELM outperforms PINN and closely matches the reference solution. As shown in Table~\ref{tab:lid-driven}, RBF-PIELM computes the solution \textbf{350× faster} in 0.4347s compared to 151.77s for PINNs, using \textbf{13.2× fewer parameters} and \textbf{1.8× fewer collocation points}, while attaining lower residual errors.
    \item \textbf{Effect of Phyics Aware Initialization} Figure~\ref{uCenterProfile} and ~\ref{vCenterProfile} compares the centerline velocities obtained from RBF-PIELM with and without Phyics Aware Initialization (PAI). We observe that RBF-PIELM with PAI outperforms RBF-PIELM without PAI and closely matches the reference solution. As shown in Table~\ref{tab:lid-driven}, RBF-PIELM with PAI has $\mathbf{17.7\%}$ \textbf{lower residual} as compared to RBF-PIELM without PAI.
\end{itemize}

\begin{table}[!htbp]
  \centering
  \begin{tabular}{lcccc}
    \toprule
    Method & Time (s) $\downarrow$ & Parameters $\downarrow$ & Collocation Pts $\downarrow$ & Residual $\downarrow$ \\
    \midrule
    PINN        & 151.77          & 9921     & 4922     & $9.84 \times 10^{-3}$ \\
    RBF-PIELM w/o PAI & \textbf{0.4260} & \textbf{750} & \textbf{2688} & $\mathbf{6.68 \times 10^{-3}}$ \\
    RBF-PIELM with PAI    & \textbf{0.4347} & \textbf{750} & \textbf{2688} & $\mathbf{5.44 \times 10^{-3}}$ \\
    \bottomrule
  \end{tabular}
    \vspace{0.5em}\caption{\centering Performance comparison for Lid-Driven Cavity. RBF-PIELM attains higher efficiency with fewer parameters and lower residual error.}
\label{tab:lid-driven}
\end{table}
\vspace{-0.4em}
\subsection{Biharmonic Equation with Oscillatory Solution}
\label{sec:test-mms}
\begin{figure}[!htbp]
    \centering
\includegraphics[width=1\linewidth]{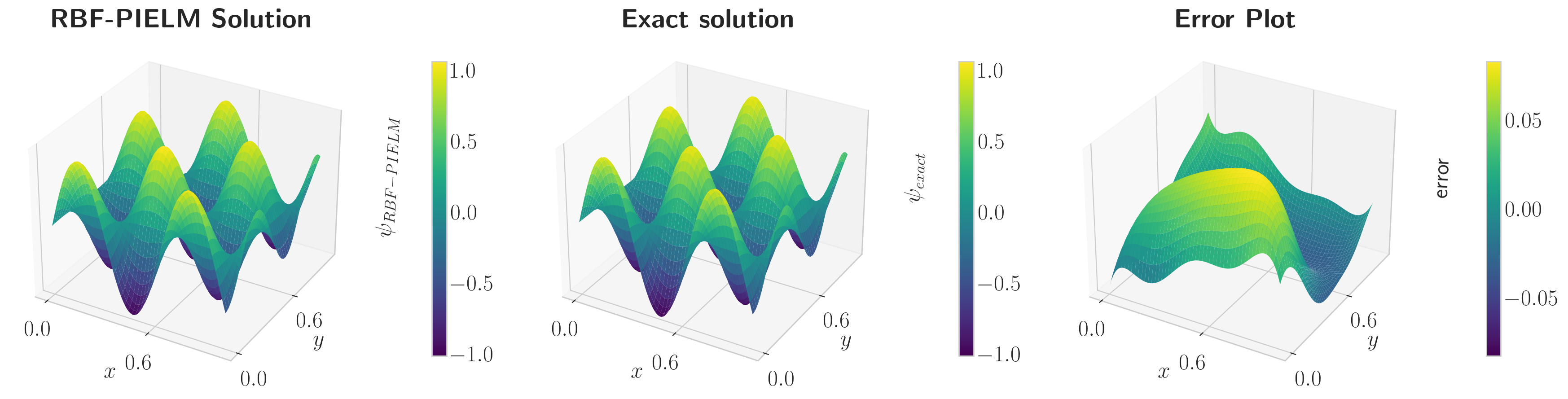}
\caption{ Manufactured biharmonic solution (\(k_1=k_2=10\)): (a) RBF-PIELM solution, (b) Exact solution and (c) Distribution of absolute error. RBF-PIELM closely matches the oscillatory exact solution with small errors.}
    \label{fig:mms3DSolution}
\end{figure}

To further assess the capability of RBF-PIELM, we consider the Method of Manufactured Solutions (MMS), which enables direct accuracy testing by prescribing an exact solution and deriving the corresponding biharmonic problem. In particular, we benchmark against Example 5.2 from~\citet{PAN2025114254}, where a fourth-order compact finite-difference scheme was evaluated. The manufactured solution involves oscillatory components (with $k_1=k_2=10$), making it a challenging test for approximation methods. We provide the derivation and exact boundary conditions imposed for the problem in Appendix Section ~\ref{mms-problem-derivation}. We sample $60 \times60$ collocation points in the domain to capture the complex oscillatory nature of the problem. Additionally, we also increase the number of RBFs to $2000$ to represent the complex solution. Unlike the lid-driven cavity case, which highlights physical relevance and efficiency, this experiment probes whether RBF-PIELM can approximate complex functions. Figure~\ref{fig:mms3DSolution} compares the exact solution, the RBF-PIELM prediction, and the error distribution. Despite the solution’s complexity, RBF-PIELM, achieves a mean error of $\mathbf{3.46\times 10^{-2}}$, thereby demonstrating its expressivity. RBF-PIELM solves the problem in 6.4 seconds with above mentioned configuration.

\vspace{-1.4em}
\section{Conclusion}
\label{sec:conclusion}
This work assessed RBF\text{-}PIELM as a lightweight, interpretable alternative to gradient-based PINNs for fourth-order PDEs. We evaluated two benchmarks: lid-driven cavity flow in a streamfunction formulation and a manufactured oscillatory solution of the biharmonic equation. On the cavity problem, RBF\text{-}PIELM achieved a \(\sim350\times\) training speedup over PINNs (0.4347\,s vs.\ 151.77\,s), used \(13.2\times\) fewer parameters, and delivered \(44.7\%\) lower error. We also showed that PAI produced better results. On the manufactured case, it reproduced the analytical field competitively, though accuracy degraded as the oscillatory content increased. These gains stem from physics-informed initialization of RBF centers (enhancing interpretability) and a single-shot least-squares solve for output weights (eliminating costly iterative training). Despite clear advantages over PINNs, RBF\text{-}PIELM still trails mature mesh-based solvers in wall-clock performance and robustness on highly oscillatory fields (Refer Appendix Section~\ref{sec:pielm-limit}), underscoring opportunities for hybrid (FEM-PIELM) formulations and residual adaptive basis refinement.

\clearpage
\bibliographystyle{unsrtnat}
\bibliography{bibliography}

\begin{thebibliography}{22}
\providecommand{\natexlab}[1]{#1}
\providecommand{\url}[1]{\texttt{#1}}
\expandafter\ifx\csname urlstyle\endcsname\relax
  \providecommand{\doi}[1]{doi: #1}\else
  \providecommand{\doi}{doi: \begingroup \urlstyle{rm}\Url}\fi

\bibitem[Strikwerda(2004)]{strikwerda2004finite}
John~C Strikwerda.
\newblock \emph{Finite difference schemes and partial differential equations}.
\newblock SIAM, 2004.

\bibitem[Hughes(2012)]{hughes2012finite}
Thomas~JR Hughes.
\newblock \emph{The finite element method: linear static and dynamic finite element analysis}.
\newblock Courier Corporation, 2012.

\bibitem[Versteeg and Malalasekera(2007)]{versteeg2007finite}
Henk~Kaarle Versteeg and Weeratunge Malalasekera.
\newblock \emph{An introduction to computational fluid dynamics: the finite volume method}.
\newblock Pearson education, 2007.

\bibitem[Yagawa(2011)]{Yagawa2011-cr}
Genki Yagawa.
\newblock Free mesh method: fundamental conception, algorithms and accuracy study.
\newblock \emph{Proc. Jpn. Acad. Ser. B Phys. Biol. Sci.}, 87\penalty0 (4):\penalty0 115--134, 2011.

\bibitem[Raissi et~al.(2019)Raissi, Perdikaris, and Karniadakis]{raissi2019physics}
Maziar Raissi, Paris Perdikaris, and George~E Karniadakis.
\newblock Physics-informed neural networks: A deep learning framework for solving forward and inverse problems involving nonlinear partial differential equations.
\newblock \emph{Journal of Computational Physics}, 378:\penalty0 686--707, 2019.

\bibitem[McGreivy and Hakim(2024)]{McGreivy2024}
Nick McGreivy and Ammar Hakim.
\newblock Weak baselines and reporting biases lead to overoptimism in machine learning for fluid-related partial differential equations.
\newblock \emph{Nature Machine Intelligence}, 6\penalty0 (10):\penalty0 1256–1269, September 2024.
\newblock ISSN 2522-5839.
\newblock \doi{10.1038/s42256-024-00897-5}.
\newblock URL \url{http://dx.doi.org/10.1038/s42256-024-00897-5}.

\bibitem[Krishnapriyan et~al.(2021)Krishnapriyan, Gholami, Zhe, Kirby, and Mahoney]{krishnapriyan2021characterizing}
Aditi Krishnapriyan, Amir Gholami, Shandian Zhe, Robert Kirby, and Michael~W Mahoney.
\newblock Characterizing possible failure modes in physics-informed neural networks.
\newblock \emph{Advances in Neural Information Processing Systems}, 34:\penalty0 26548--26560, 2021.

\bibitem[Dwivedi and Srinivasan(2019)]{dwivedi2019physics}
Vikas Dwivedi and Balaji Srinivasan.
\newblock Physics informed extreme learning machine (pielm)--a rapid method for the numerical solution of partial differential equations.
\newblock \emph{arXiv preprint arXiv:1907.03507}, 2019.

\bibitem[Huang et~al.(2006{\natexlab{a}})Huang, Zhu, and Siew]{huang2006extreme}
Guang-Bin Huang, Qin-Yu Zhu, and Chee-Kheong Siew.
\newblock Extreme learning machine: Theory and applications.
\newblock \emph{Neurocomputing}, 70\penalty0 (1-3):\penalty0 489--501, 2006{\natexlab{a}}.

\bibitem[Shalev-Shwartz et~al.(2017)Shalev-Shwartz, Shamir, and Shammah]{shalev2017failures}
Shai Shalev-Shwartz, Ohad Shamir, and Shaked Shammah.
\newblock Failures of gradient-based deep learning.
\newblock \emph{arXiv preprint arXiv:1703.07950}, 2017.
\newblock URL \url{https://arxiv.org/abs/1703.07950}.

\bibitem[Huang et~al.(2006{\natexlab{b}})Huang, Chen, and Siew]{huang2006universal}
Guang-Bin Huang, Lei Chen, and Chai~Quek Siew.
\newblock Universal approximation using incremental constructive feedforward networks with random hidden nodes.
\newblock \emph{IEEE Transactions on Neural Networks}, 17\penalty0 (4):\penalty0 879--892, 2006{\natexlab{b}}.
\newblock \doi{10.1109/TNN.2006.875977}.

\bibitem[Dwivedi and Srinivasan(2020)]{dwivedi2020physics}
Vikas Dwivedi and Balaji Srinivasan.
\newblock Physics informed extreme learning machine (pielm)--a rapid method for the numerical solution of partial differential equations.
\newblock \emph{Neurocomputing}, 391:\penalty0 96--118, 2020.

\bibitem[Xu et~al.(2022)]{xu2022bayesian}
Liu Xu et~al.
\newblock Bayesian physics-informed extreme learning machine for forward and inverse pde problems with noisy data.
\newblock \emph{arXiv preprint arXiv:2205.06948}, 2022.

\bibitem[Dwivedi et~al.(2025{\natexlab{a}})Dwivedi, Sixou, and Sigovan]{DWIVEDI2025130924}
Vikas Dwivedi, Bruno Sixou, and Monica Sigovan.
\newblock Curriculum learning-driven pielms for fluid flow simulations.
\newblock \emph{Neurocomputing}, 650:\penalty0 130924, 2025{\natexlab{a}}.
\newblock ISSN 0925-2312.
\newblock \doi{https://doi.org/10.1016/j.neucom.2025.130924}.
\newblock URL \url{https://www.sciencedirect.com/science/article/pii/S0925231225015966}.

\bibitem[Buhmann(2003)]{buhmann2003radial}
Martin~D. Buhmann.
\newblock \emph{Radial Basis Functions: Theory and Implementations}.
\newblock Cambridge University Press, Cambridge, 2003.
\newblock ISBN 9780521101332.

\bibitem[Schaback(2006)]{schaback2007kernel}
Robert Schaback.
\newblock Kernel techniques: from machine learning to meshless methods.
\newblock \emph{Acta Numerica}, 15:\penalty0 543--639, 2006.

\bibitem[Kansa(1990)]{kansa1990multiquadrics}
EJ~Kansa.
\newblock Multiquadrics—a scattered data approximation scheme with applications to computational fluid-dynamics—i surface approximations and partial derivative estimates.
\newblock \emph{Computers \& Mathematics with Applications}, 19\penalty0 (8-9):\penalty0 127--145, 1990.

\bibitem[Park and Sandberg(1991)]{Park1991}
J.~Park and I.~W. Sandberg.
\newblock Universal approximation using radial-basis-function networks.
\newblock \emph{Neural Computation}, 3\penalty0 (2):\penalty0 246–257, June 1991.
\newblock ISSN 1530-888X.
\newblock \doi{10.1162/neco.1991.3.2.246}.
\newblock URL \url{http://dx.doi.org/10.1162/neco.1991.3.2.246}.

\bibitem[Pan et~al.(2025)Pan, Li, Li, and Hu]{PAN2025114254}
Kejia Pan, Jin Li, Zhilin Li, and Hongling Hu.
\newblock A fourth order mixed compact finite difference scheme for biharmonic equations.
\newblock \emph{Journal of Computational Physics}, 539:\penalty0 114254, 2025.
\newblock ISSN 0021-9991.
\newblock \doi{https://doi.org/10.1016/j.jcp.2025.114254}.
\newblock URL \url{https://www.sciencedirect.com/science/article/pii/S0021999125005376}.

\bibitem[Marchi et~al.(2009)Marchi, Suero, and Araki]{marchietal}
Carlos Marchi, Roberta Suero, and Luciano Araki.
\newblock The lid-driven square cavity flow: Numerical solution with a 1024 x 1024 grid.
\newblock \emph{Journal of The Brazilian Society of Mechanical Sciences and Engineering - J BRAZ SOC MECH SCI ENG}, 31, 07 2009.
\newblock \doi{10.1590/S1678-58782009000300004}.

\bibitem[Dwivedi et~al.(2025{\natexlab{b}})Dwivedi, Srinivasan, Sigovan, and Sixou]{dwivedi2025kerneladaptivepielmsforwardinverse}
Vikas Dwivedi, Balaji Srinivasan, Monica Sigovan, and Bruno Sixou.
\newblock Kernel-adaptive pi-elms for forward and inverse problems in pdes with sharp gradients, 2025{\natexlab{b}}.
\newblock URL \url{https://arxiv.org/abs/2507.10241}.

\bibitem[Radice(2021)]{radice2021airy_biharmonic}
J.~Radice, J.\.
\newblock A biharmonic polynomial airy stress function for the square-end adhesive layer and sandwich core.
\newblock \emph{Mechanics Research Communications}, 113:\penalty0 103684, 2021.

\end{thebibliography}

\appendix
\section{Boundary Conditions for Lid-Driven Cavity}
\label{sec:BCLidDriven}
\begin{figure}[h!]
    \centering
    \includegraphics[width=0.6\textwidth]{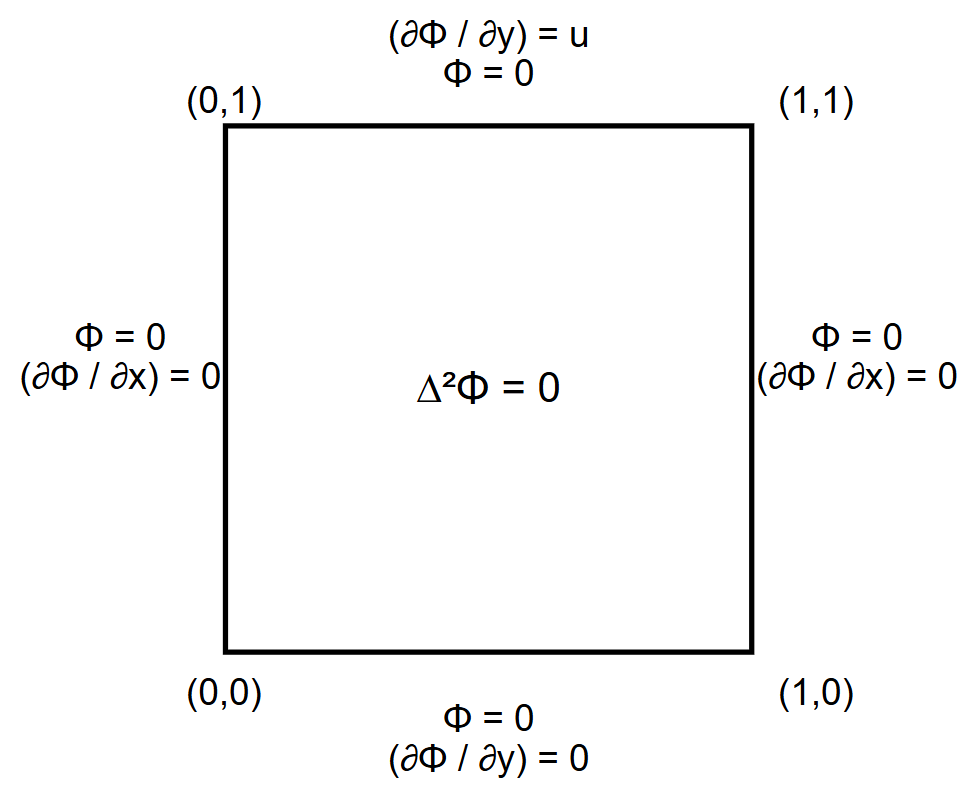}
 \caption{Lid-driven cavity benchmark: moving top wall (\(u=1\)) and stationary side/bottom walls.}
    \label{fig:lid_cavity}
\end{figure}

The boundary condition imposed are as depicted in Figure~\ref{fig:lid_cavity}.

\section{Chebyshev Spacing for Sampling and Kernel Placement}
\label{sec:chebyshev}

In our experiments, collocation points are distributed using Chebyshev spacing along each coordinate axis. For $N$ points in the interval $[0,1]$, the Chebyshev nodes are defined as
\begin{equation}
x_j = \tfrac{1}{2}\left[1 - \cos\left(\tfrac{j\pi}{N-1}\right)\right], \quad j=0,1,\dots,N-1.
\end{equation}

\begin{figure}[!htbp]
    \centering
\subfloat[Distribution of collocation and boundary points.]{%
        \label{fig:chebyshevSpacing}\includegraphics[width=0.48\textwidth]{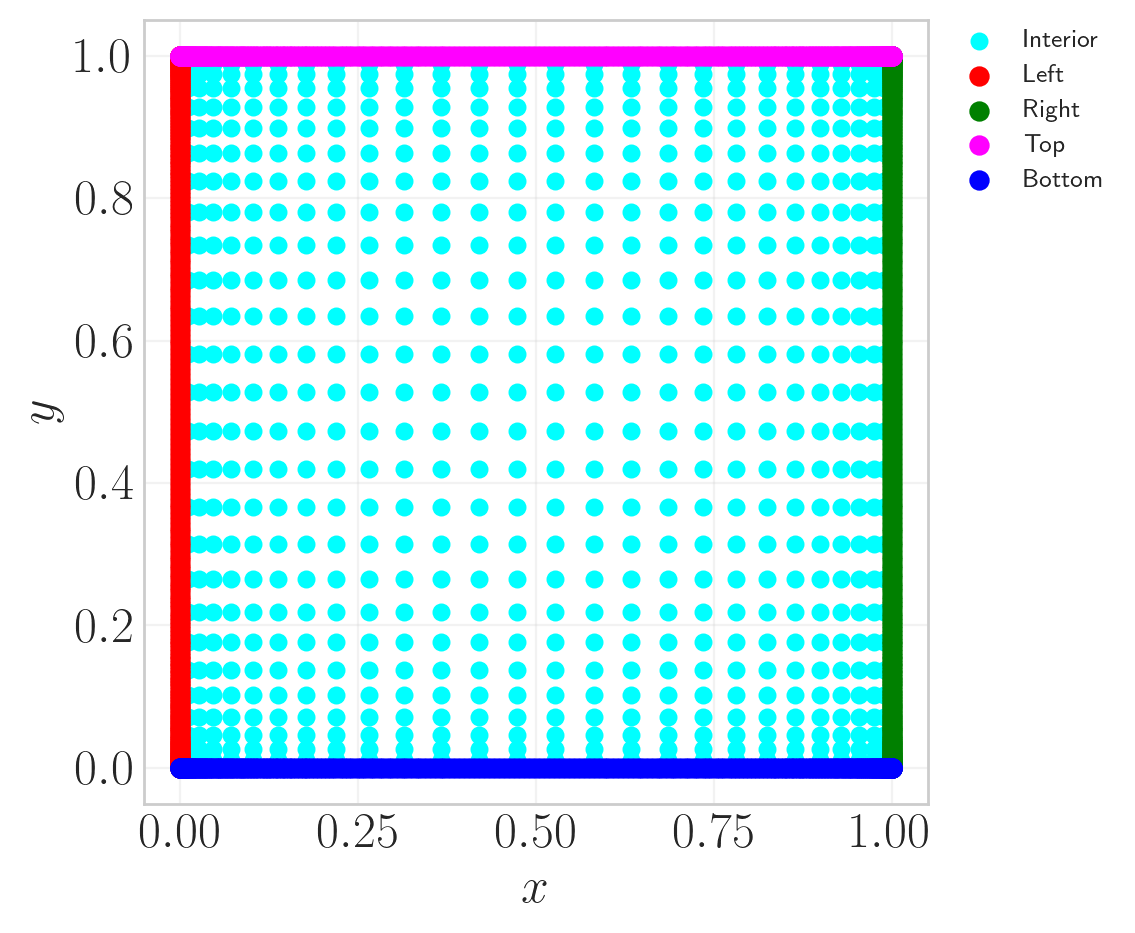}
    }
    \hfill
    \subfloat[Visualization of RBF kernels. RBF centers (red) with widths encoded by the radii of blue disks. ]{%
        \label{fig:rbfcenters}\includegraphics[width=0.48\textwidth]{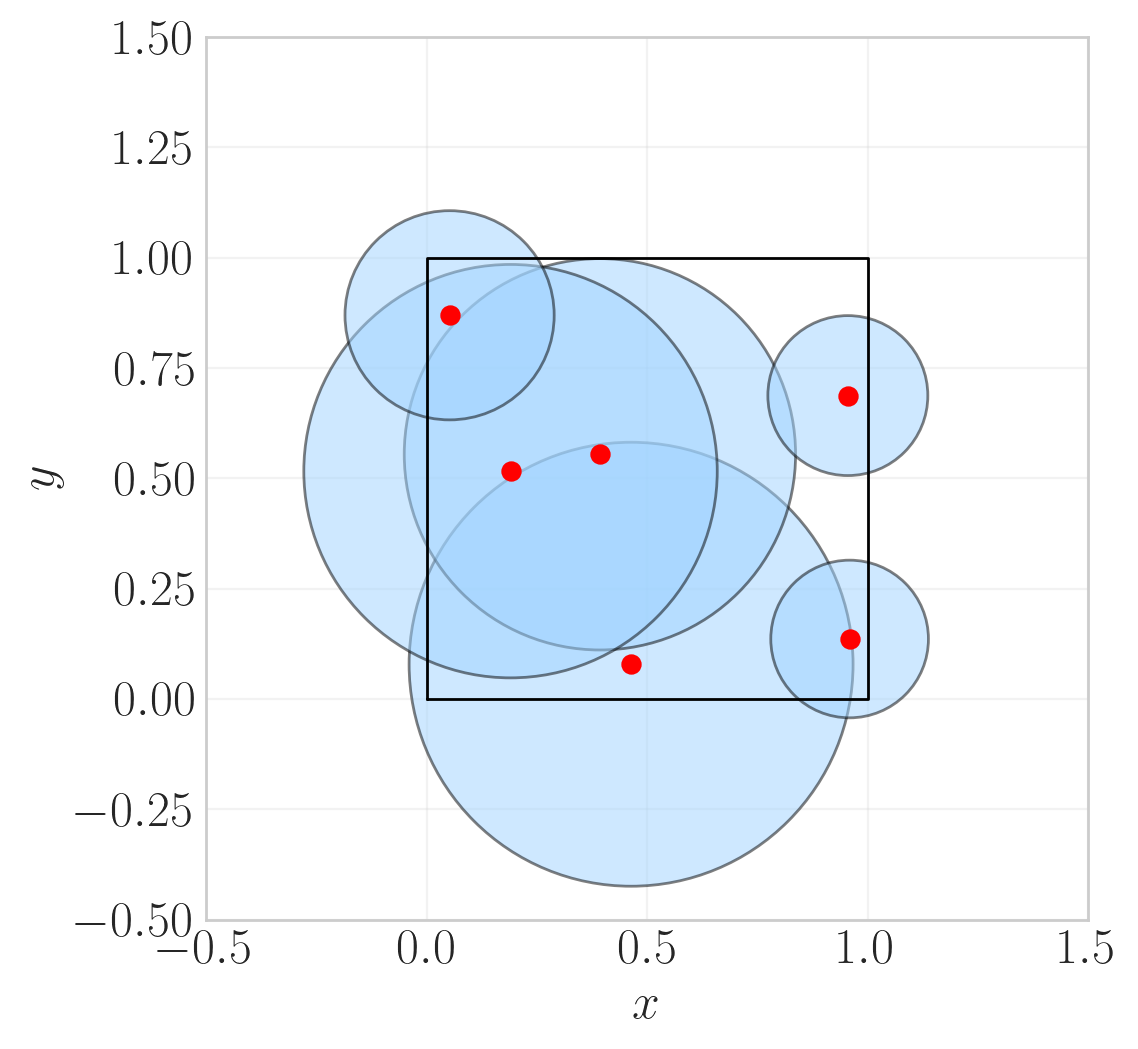}
    }
    \caption{Collocation/boundary points and RBF width visualization for Test Case 1.}
    \label{fig:pielm-flow-results}
\end{figure}

This distribution concentrates points near domain boundaries while spacing them more sparsely in the interior. Such clustering is advantageous in the lid-driven cavity problem, where strong velocity gradients occur near walls and uniform sampling may under-resolve them. Chebyshev nodes are also well established in spectral and collocation methods for reducing interpolation error and improving stability. For mesh-free solvers like RBF-PIELM, this spacing provides boundary refinement without ad hoc clustering, balancing interior coverage with boundary resolution and leading to more efficient, accurate solutions (Figure~\ref{fig:chebyshevSpacing}).

\section{Manufactured Solution Details for Example 5.2 from \citeauthor{PAN2025114254}}
\label{mms-problem-derivation}

In Section~\ref{sec:test-mms}, the method of manufactured solutions is employed from \citeauthor{PAN2025114254} to rigorously validate the accuracy of our numerical solver for biharmonic equations. Specifically, we select an exact, oscillatory solution of the form
\begin{equation}
    u(x, y) = \sin(k_1 x) \cos(k_2 y), \quad (x, y) \in (0,1)^2,
\end{equation}
where $k_1$ and $k_2$ are prescribed parameters (here, both set to 10).

To construct the corresponding source term $f(x,y)$ required for the biharmonic equation
\begin{equation}
    \Delta^2 u(x, y) = f(x, y),
\end{equation}
we substitute the analytic solution $u(x, y)$ into the biharmonic operator. In two dimensions, the biharmonic operator is given by
\begin{equation}
    \Delta^2 u = \frac{\partial^4 u}{\partial x^4} + 2 \frac{\partial^4 u}{\partial x^2 \partial y^2} + \frac{\partial^4 u}{\partial y^4}.
\end{equation}
Applying this operator to $u(x, y)$,
\begin{align}
    \frac{\partial^2 u}{\partial x^2} &= -k_1^2 \sin(k_1 x) \cos(k_2 y), \\
    \frac{\partial^2 u}{\partial y^2} &= -k_2^2 \sin(k_1 x) \cos(k_2 y),
\end{align}
and continuing to fourth derivatives,
\begin{align}
    \frac{\partial^4 u}{\partial x^4} &= k_1^4 \sin(k_1 x) \cos(k_2 y), \\
    \frac{\partial^4 u}{\partial y^4} &= k_2^4 \sin(k_1 x) \cos(k_2 y), \\
    \frac{\partial^4 u}{\partial x^2 \partial y^2} &= k_1^2 k_2^2 \sin(k_1 x) \cos(k_2 y),
\end{align}
which yields
\begin{align}
    f(x, y) &= \Delta^2 u(x, y) \\
            &= [k_1^4 + 2k_1^2 k_2^2 + k_2^4]\sin(k_1 x) \cos(k_2 y).
\end{align}

This manufactured source term $f(x, y)$ is used in the numerical experiments to assess the solver's accuracy, with the exact solution available for direct comparison. 
\paragraph{Boundary Conditions} The boundary conditions imposed are Dirchelet in nature and are obatined from function values.

\section{Limitations of RBF-PIELM}
\label{sec:pielm-limit}
\begin{figure}[!htbp]
    \centering
    \includegraphics[width=1\linewidth]{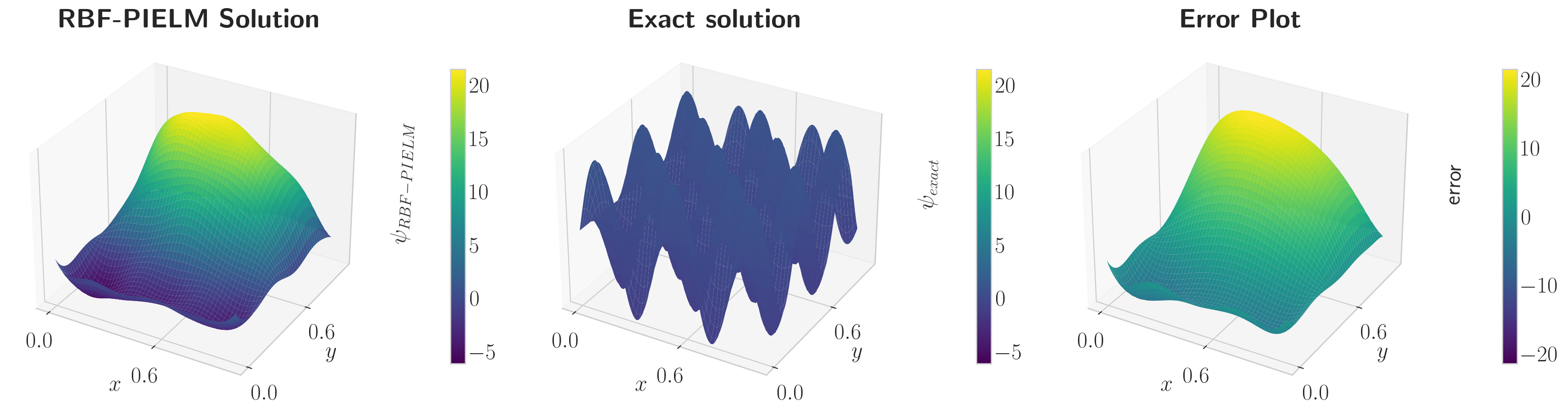}
    \caption{Manufactured biharmonic solution (\(k_1=k_2=20\)): (a) RBF-PIELM solution, (b) Exact solution, and (c) Distribution of absolute error. RBF-PIELM breaks down on this highly oscillatory case.}
    \label{fig:pielm-limit}
\end{figure}

We here show a more complex solution from~\cite{PAN2025114254} by setting $k_1=k_2=20$. We present the result in Figure~\ref{fig:pielm-limit}. We observe that RBF-PIELM faces several challenges in representing the solution more granularly. 

\section{Additional Results for Test Case 1 }



\begin{figure}[!htbp]
\centering
\subfloat[ \centering Streamfunction \(\psi\).\label{psiPlotPIELM}]{%
    \includegraphics[width=0.51\textwidth]{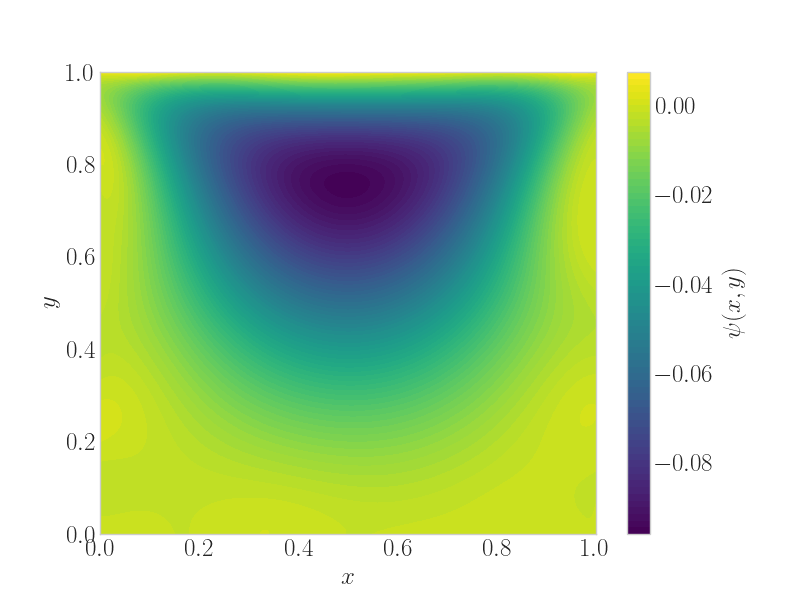}
}
\hfill
\subfloat[\centering Velocity magnitude \(\lVert \mathbf{u}\rVert\).\label{velMagPlotPIELM}]{%
    \includegraphics[width=0.46\textwidth]{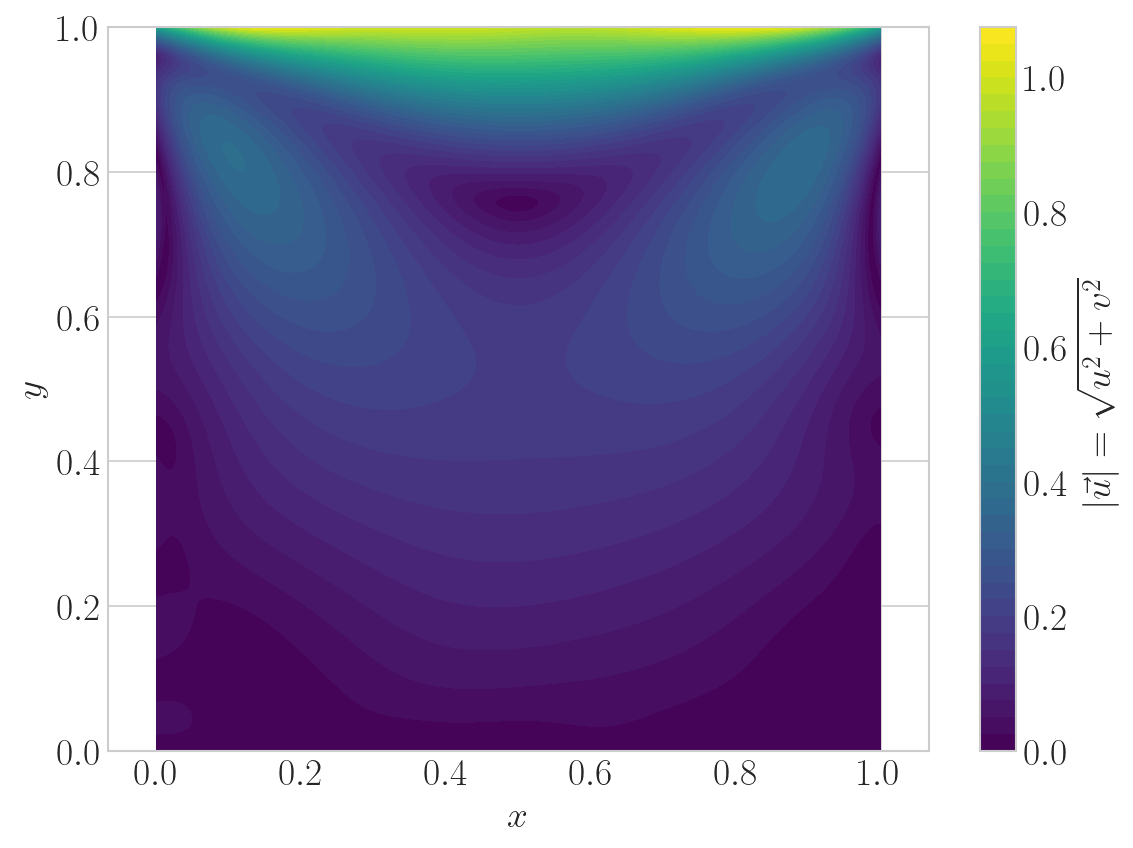}
}
\caption{RBF-PIELM predictions of streamfunction \(\psi\) and velocity magnitude field \(\lVert \mathbf{u}\rVert\).}
\label{fig:pielm_fields}

\end{figure}

\section{RBF Width Tuning}
\label{sec:hyper-param}
\begin{figure}[!htbp]
\centering
\subfloat[\centering Residual contours over \((\sigma_0,\sigma_c)\).\label{fig:hyperparamplot}]{%
    \includegraphics[width=0.48\textwidth]{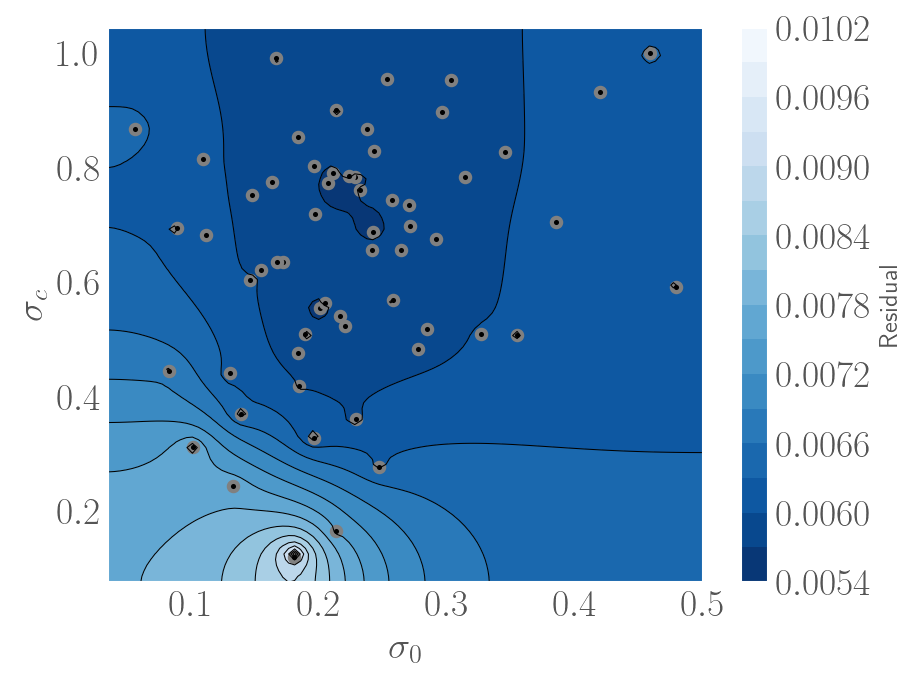}
}
\hfill
\subfloat[\centering Residual as a function of \(\sigma_0\) (with \(\sigma_c\) fixed).\label{fig:hyperparamplot2}]{%
    \includegraphics[width=0.48\textwidth]{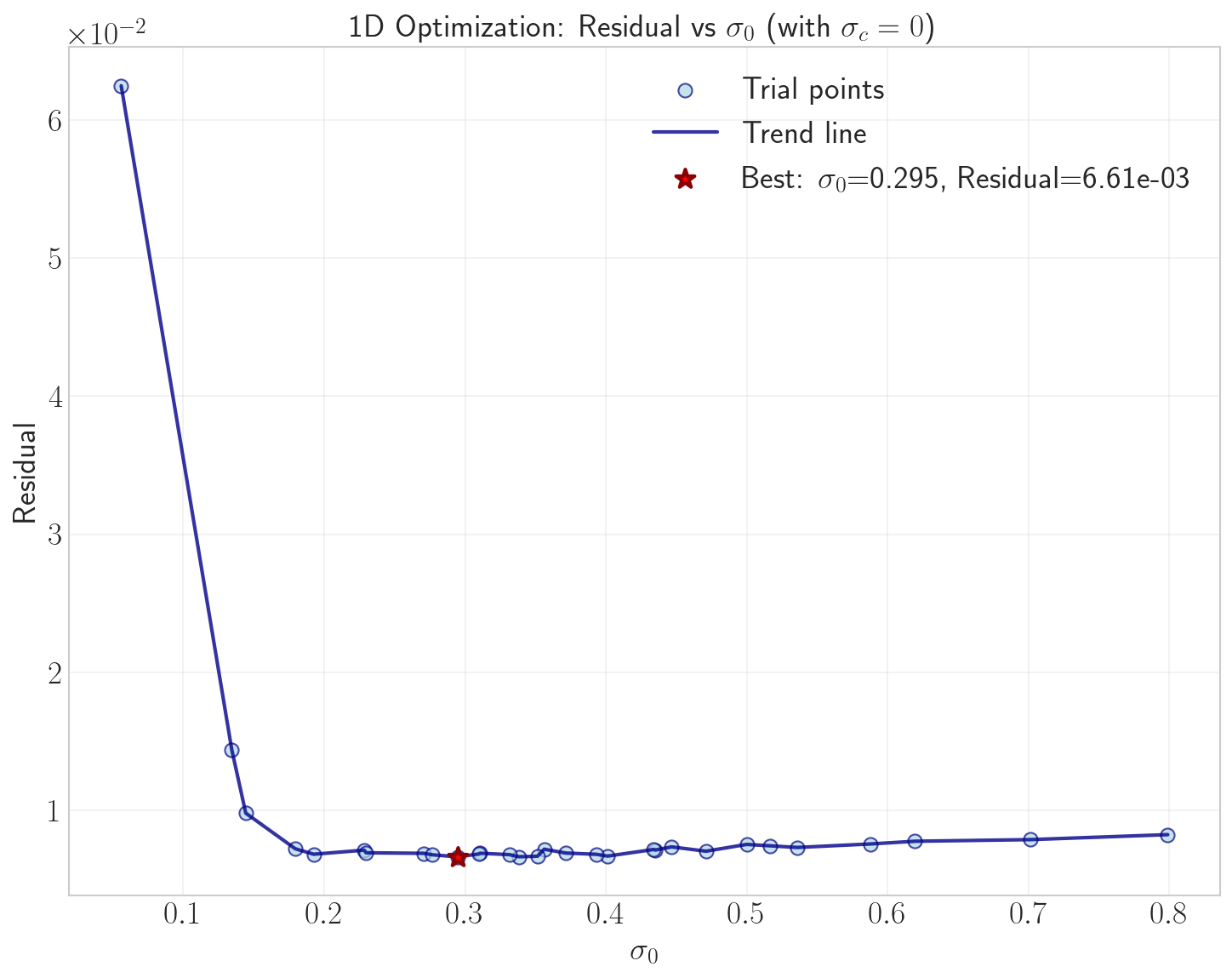}
}
\subfloat[\centering Residual as a function of Number of Neurons ($N^*$).\label{fig:hyperparamplot3}]{%
    \includegraphics[width=0.48\textwidth]{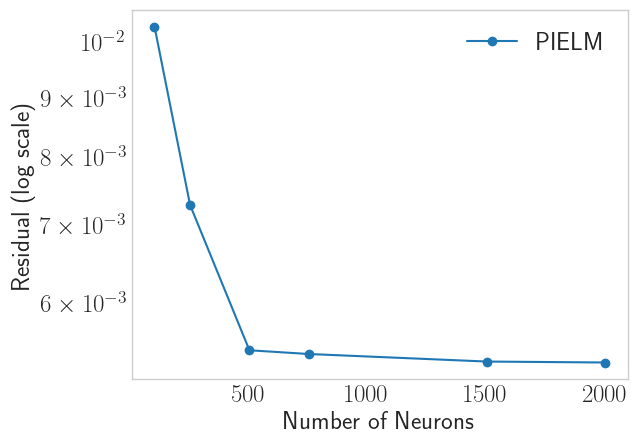}
}
\caption{\centering Hyperparameter sensitivity of the residual: tuning \(\sigma_0\) and \(\sigma_c\) reduces the residual by approximately \(50\%\).}
\label{fig:hyperparam}

\end{figure}

To get the best performance from RBF-PIELM we tune the hyperparameters of $\sigma$ function, namely, $\sigma_0$ and $\sigma_c$. Figure~\ref{fig:hyperparamplot} shows the contours of residuals with x and y axis being different axes of $\sigma_0$ and $\sigma_c$ respectively. We see that tuning the hyperparameters decreases the residual by around $50\%$. We also show the variations of Residual with number of Neurons/RBFs in Figure~\ref{fig:hyperparamplot3}. We see that as the number of neurons increases, at a certain point the residual saturates.


\section{Neural Network Specifications}
\label{nn-specs}
For training PINNs, we use the PyTorch\footnote{https://pytorch.org/} library with Nvidia Tesla P100, offered through \href{kaggle.com}{Kaggle}. Further the network used has 5 layers with the following number of neurons: $[2, 40, 120, 40, 1]$. The optimizer used was Adam with learning rate set to $5\times10^{-3}$.
\section{Compute Specifications}
\label{sec:compute-specs}
For RBF-PIELM Experiments, we run the experiment only on a CPU namely, a Ryzen 7 5800H with 16GB RAM with a single process. For PINN based experiments, we use Nvidia Tesla P100, offered through \href{kaggle.com}{Kaggle}.
\newpage
\section*{NeurIPS Paper Checklist}



\begin{enumerate}

\item {\bf Claims}
    \item[] Question: Do the main claims made in the abstract and introduction accurately reflect the paper's contributions and scope?
    \item[] Answer: \answerYes{} 
    \item[] Justification: The main claims and contributions made in the manuscript are articulated in abstract line numbers 9-15 and Introduction line numbers 48-55.

\item {\bf Limitations}
    \item[] Question: Does the paper discuss the limitations of the work performed by the authors?
    \item[] Answer: \answerYes{} 
    \item[] Justification: Please refer to Section~\ref{sec:conclusion}, line numbers 130-133 for the limitations and Future Work of RBF-PIELM.

\item {\bf Theory assumptions and proofs}
    \item[] Question: For each theoretical result, does the paper provide the full set of assumptions and a complete (and correct) proof?
    \item[] Answer: \answerYes{} 
    \item[] Justification: Details of the Method of Manufactured Solutions are presented in Section~\ref{sec:test-mms}, with the derivation outlined in Appendix Section~\ref{mms-problem-derivation}. The mathematical formulation of the proposed RBF-PIELM framework is provided in Section~\ref{sec:rbf-pielm-math}.

    \item {\bf Experimental result reproducibility}
    \item[] Question: Does the paper fully disclose all the information needed to reproduce the main experimental results of the paper to the extent that it affects the main claims and/or conclusions of the paper (regardless of whether the code and data are provided or not)?
    \item[] Answer: \answerYes{} 
    \item[] Justification: We provide the Formulation to reproduce RBF-PIELM with and without Physics Aware Initialization in Sections~\ref{sec:rbf-pielm-math} and~\ref{sec:test-lid-driven}. We also provide the sampling method for collocation points and RBF Centers in Sections~\ref{sec:test-lid-driven}, ~\ref{sec:chebyshev} and ~\ref{sec:hyper-param}. Additionally, we also provide the code to reproduce all the results with documentation on its usage.

\item {\bf Open access to data and code}
    \item[] Question: Does the paper provide open access to the data and code, with sufficient instructions to faithfully reproduce the main experimental results, as described in supplemental material?
    \item[] Answer: \answerYes{} 
    \item[] Justification: We provide the anonymised code with documentation and instructions to reproduce the results as a anonymised repository. 
    \item[] Guidelines:
    \begin{itemize}
        \item The answer NA means that paper does not include experiments requiring code. No datasets are required to reproduce the results.
        \item Please see the NeurIPS code and data submission guidelines (\url{https://nips.cc/public/guides/CodeSubmissionPolicy}) for more details.
        \item While we encourage the release of code and data, we understand that this might not be possible, so “No” is an acceptable answer. Papers cannot be rejected simply for not including code, unless this is central to the contribution (e.g., for a new open-source benchmark).
        \item The instructions should contain the exact command and environment needed to run to reproduce the results. See the NeurIPS code and data submission guidelines (\url{https://nips.cc/public/guides/CodeSubmissionPolicy}) for more details.
        \item The authors should provide instructions on data access and preparation, including how to access the raw data, preprocessed data, intermediate data, and generated data, etc.
        \item The authors should provide scripts to reproduce all experimental results for the new proposed method and baselines. If only a subset of experiments are reproducible, they should state which ones are omitted from the script and why.
        \item At submission time, to preserve anonymity, the authors should release anonymized versions (if applicable).
        \item Providing as much information as possible in supplemental material (appended to the paper) is recommended, but including URLs to data and code is permitted.
    \end{itemize}

\item {\bf Experimental setting/details}
    \item[] Question: Does the paper specify all the training and test details (e.g., data splits, hyperparameters, how they were chosen, type of optimizer, etc.) necessary to understand the results?
    \item[] Answer: \answerYes{} 
    \item[] Justification:  The sampling strategies for collocation points and RBF centers are described in Sections~\ref{sec:test-lid-driven} and~\ref{sec:chebyshev}. All linear algebra operations are implemented using standard libraries such as NumPy. The experimental settings for training PINNs are detailed in Appendix~\ref{nn-specs}.

\item {\bf Experiment statistical significance}
    \item[] Question: Does the paper report error bars suitably and correctly defined or other appropriate information about the statistical significance of the experiments?
    \item[] Answer: \answerYes{} 
    \item[] Justification: 
    To assess robustness, we first analyze the effect of kernel width parameters $\sigma_0$ and $\sigma_c$
 (Section~\ref{sec:hyper-param}), followed by experiments with different numbers of neurons. The corresponding error plots quantify sensitivity to hyperparameters and model capacity. Since our Physics-Informed Extreme Learning Machine does not rely on training data, conventional train–test splits and data-driven error bars are not directly applicable. The only source of randomness lies in the sampling of RBF centers, making the observed error variation small; thus, reliability is demonstrated through consistent results across repeated runs under varying settings.

\item {\bf Experiments compute resources}
    \item[] Question: For each experiment, does the paper provide sufficient information on the computer resources (type of compute workers, memory, time of execution) needed to reproduce the experiments?
    \item[] Answer: \answerYes{} 
    \item[] Justification: The compute specifications are provided in Appendix Section~\ref{sec:compute-specs}.
    
\item {\bf Code of ethics}
    \item[] Question: Does the research conducted in the paper conform, in every respect, with the NeurIPS Code of Ethics \url{https://neurips.cc/public/EthicsGuidelines}?
    \item[] Answer: \answerYes{} 
    \item[] Justification: We have reviewed the NeurIPS Code of Ethics and we acknowledge that we conform to it.

\item {\bf Broader impacts}
    \item[] Question: Does the paper discuss both potential positive societal impacts and negative societal impacts of the work performed?
    \item[] Answer: \answerNA{} 
    \item[] Justification: The work is primarily methodological and technical in nature, with no direct or immediate societal impact
    
\item {\bf Safeguards}
    \item[] Question: Does the paper describe safeguards that have been put in place for responsible release of data or models that have a high risk for misuse (e.g., pretrained language models, image generators, or scraped datasets)?
    \item[] Answer: \answerNA{} 
    \item[] Justification: The paper does not involve the release of data or models with potential for misuse; hence, no specific safeguards are required.

\item {\bf Licenses for existing assets}
    \item[] Question: Are the creators or original owners of assets (e.g., code, data, models), used in the paper, properly credited and are the license and terms of use explicitly mentioned and properly respected?
    \item[] Answer: \answerYes{} 
    \item[] Justification: All external assets used in this work are properly credited through citations to their respective publications and code repositories, with licenses and terms of use duly respected.

\item {\bf New assets}
    \item[] Question: Are new assets introduced in the paper well documented and is the documentation provided alongside the assets?
    \item[] Answer: \answerYes{} 
    \item[] Justification: The code released with the paper is accompanied by clear documentation to facilitate reproducibility and ease of use.

\item {\bf Crowdsourcing and research with human subjects}
    \item[] Question: For crowdsourcing experiments and research with human subjects, does the paper include the full text of instructions given to participants and screenshots, if applicable, as well as details about compensation (if any)? 
    \item[] Answer: \answerNA{} 
    \item[] Justification: Not applicable, as the work does not involve crowd-sourcing experiments or research with human subjects.

\item {\bf Institutional review board (IRB) approvals or equivalent for research with human subjects}
    \item[] Question: Does the paper describe potential risks incurred by study participants, whether such risks were disclosed to the subjects, and whether Institutional Review Board (IRB) approvals (or an equivalent approval/review based on the requirements of your country or institution) were obtained?
    \item[] Answer: \answerNA{} 
    \item[] Justification: Not applicable, as the work does not involve human subjects and therefore does not require IRB approval.

\item {\bf Declaration of LLM usage}
    \item[] Question: Does the paper describe the usage of LLMs if it is an important, original, or non-standard component of the core methods in this research? Note that if the LLM is used only for writing, editing, or formatting purposes and does not impact the core methodology, scientific rigorousness, or originality of the research, declaration is not required.
    \item[] Answer: \answerNA{} 
    \item[] Justification: Not applicable, as large language models were not used in the core methodology or scientific contributions of this work.

\end{enumerate}

\end{document}